\documentclass[conference]{IEEEtran}
\IEEEoverridecommandlockouts
\usepackage{cite}
\usepackage{adjustbox}
\usepackage{amsmath,amssymb,amsfonts}
\usepackage{algorithmic}
\usepackage{graphicx}
\usepackage{textcomp}
\usepackage{hyperref}
\usepackage{xcolor}
\def\BibTeX{{\rm B\kern-.05em{\sc i\kern-.025em b}\kern-.08em
    T\kern-.1667em\lower.7ex\hbox{E}\kern-.125emX}}
\begin{document}

\title{Using Machine Intelligence to Prioritise Code Review Requests}

\author{\IEEEauthorblockN{Nishrith Saini}
\IEEEauthorblockA{\textit{Ericsson AB, Sweden.} \\
nishrith.saini@ericsson.com}
\and
\IEEEauthorblockN{Ricardo Britto}
\IEEEauthorblockA{\textit{Ericsson AB, Sweden.} \\
\textit{Blekinge Institute of Technology, Sweden.}\\
ricardo.britto@ericsson.com}
}
\maketitle

\begin{abstract}
Modern Code Review (MCR) is the process of reviewing new code changes that need to be merged with an existing codebase. As a developer, one may receive many code review requests every day, i.e., the review requests need to be prioritised. Manually prioritising review requests is a challenging and time-consuming process. To address the above problem, we conducted an industrial case study at Ericsson aiming at developing a tool called Pineapple, which uses a Bayesian Network to prioritise code review requests. To validate our approach/tool, we deployed it in a live software development project at Ericsson, wherein more than 150 developers develop a telecommunication product. We focused on evaluating the predictive performance, feasibility, and usefulness of our approach. The results indicate that Pineapple has competent predictive performance (RMSE = 0.21 and MAE = 0.15). Furthermore, around 82.6\% of Pineapple's users believe the tool can support code review request prioritisation by providing reliable results, and around 56.5\% of the users believe it helps reducing code review lead time. As future work, we plan to evaluate Pineapple's predictive performance, usefulness, and feasibility through a longitudinal investigation.
\end{abstract}


\section{Introduction}
Software code review is the practice that involves team members to check/critique the changes made to an existing software system before the code changes are integrated into the target central code base \cite{shimagaki2016study}. Code reviews help in improving the quality of the code and reduce post-integration defects \cite{ackerman1984software, aurum2002state, fagan2002design}. Performing code reviews has been an essential practice in software development \cite{mcintosh2014impact, ouni2016search}. The traditional code review processes were cumbersome and time-consuming \cite{shull2008inspecting}, which led to the introduction of modern code review.

Modern Code Review (MCR) is a lightweight alternative to the traditional code review process and it is practised widely in open source software projects and large organisations \cite{bacchelli2013expectations}. Dedicated tools such as Gerrit and Github are used to make the modern code review process manageable, comfortable and effective \cite{ouni2016search, beller2014modern}.

Incorporating MCR in the software development process helps in producing software with fewer bugs and easy maintenance \cite{ouni2016search}. MCR allows developers to monitor code changes and give feedback before new code changes are integrated into a project's code base \cite{ram2018makes}. MCR also helps to understand code and share knowledge \cite{mcintosh2014impact, ouni2016search, bacchelli2013expectations}.

As a developer, one may receive many code review requests (also known as pull requests or merge requests) every day. Prioritising these requests is considered as one of the biggest concern in their day-to-day work \cite{gousios2015work}. 

Having received a large number of review requests, a reviewer has to select the requests manually to start reviewing the associated code changes. In general, the prioritisation and selection are made according to the time the review requests are created rather than their relevance. Sometimes, the review requests are not prioritised at all \cite{gousios2015work}. 

Code review requests are often sent to multiple reviewers. It may be the case that one reviewer has already reviewed a code change, i.e., it is no longer necessary to review the same code change again. Another important aspect is that some of the code review requests may be minor changes to the code base, i.e., they do not need the utmost attention of the reviewers. 

The aspects mentioned above make the traditional manual process of prioritising and selecting review requests time consuming and inefficient. Implications of this manual process are as follows:
\begin{itemize}
    \item[$\cdot$] The time spent prioritising review requests costs not only development time but also increases the overall lead time of the code review process \cite{bacchelli2013expectations}. 
    \item[$\cdot$] A small-time delay in the review process may seem negligible, but as they occur several times, it affects the overall performance and development speed of an individual or a team.
\end{itemize}

One option to improve how MCR is done is to prioritise code review requests automatically. Automating the prioritisation of code review requests may help to decrease the lead time of a code review process, reduce the workload for the reviewers, and increase the development efficiency.

In this paper, we report the findings from developing, deploying, using, and evaluating a tool called Pineapple, which automates the code review prioritisation activity. Our main goal was to introduce a machine intelligence-based (AI/Machine Learning) tool that helps code reviewers to prioritise code review requests that, in turn, helps in decreasing the overall lead time of code review process along with helping to reduce the workload of the reviewers. Existing literature shows that Machine Intelligence techniques (e.g., Bayesian Networks\cite{friedman1997bayesian}) have been used to solve prioritisation problems in other domains \cite{yoo2009clustering, busetta2017tool, but2007performance}, which is the reason we have selected this type of technique. 

We answered the following research questions in this paper:
\begin{itemize}
    \item[$\cdot$] RQ1~--~What are the factors that relate to prioritising code review requests in an effective way?
    \\
    \item[$\cdot$] RQ2~--~What is the predictive performance of Pineapple?
    \\
    \item[$\cdot$] RQ3~--~How feasible and useful is Pineapple in a large-scale live industrial environment?
    \\
\end{itemize}

The main contributions of this paper are as follows:
\begin{itemize}
    \item[$\cdot$] The identification of factors (and the relationship between them) that relate to effective prioritisation of code review requests.
    \item[$\cdot$] A machine intelligence-based tool (Pineapple) that uses a Bayesian Network.
    \item[$\cdot$] An empirical evaluation of Pineapple in an industrial environment that focuses on its predictive performance, usefulness, and feasibility.
\end{itemize}

\section{Related Work}
\label{chp:relatedwork}

With the popularity of Modern Code Reviews, many efforts have been made to improve this process. In the efforts made to reach this goal, many challenges have been found.

Reviewers need to have an in-depth understanding and experience related to the changes they are working with \cite{beller2014modern, thongtanunam2014improving}. If a reviewer does not have enough knowledge, it makes it difficult for him/her to review the changes. It also makes it difficult for him/her to prioritise review request among the other review requests he/she received, which consumes time.

Gousios \textit{et al}. \cite{gousios2015work} have conducted a survey that focused on the challenges faced by integrators in open source projects. This survey helped in identifying that prioritisation of code review requests is one of the major tasks that concern software developers. The authors have also identified factors that code reviewers consider while prioritising review requests. The main identified factors are the size of the change, complexity, age of the change, merge conflicts, and type of the change. 

Some of the related papers focus on proposing models to predict if a code change will be merged or abandoned rather then producing a prioritised list of code review requests. They also have another common aspect: the proposed approaches are only evaluated from a predictive performance point of view using historical data from open source projects. Jeong \textit{et al}. \cite{jeong2009improving} have identified a set of features that help to find out if the code in a given review request will be merged or abandoned. They implemented a Weka-based Bayesian Network. They focused on a set of keywords in code changes that are extracted from bug reports. Similarly, Gousios \textit{et al}. \cite{gousios2014exploratory} have identified a set of 12 features that help in predicting if a review request will get merged and used the identified factors to implement a Random Forest model. Similar research, but with a different approach, was done by Zhao \textit{et al}.  \cite{zhao2019improving}. They have proposed learn-to-rank models to predict whether the pull request will be merged or abandoned.

Fan \textit{et al}. \cite{fan2018early} focused on predict if a code change associated with a code review request will be accepted or not. The authors used historical data from open source projects to evaluate their approach. They also used Random Forest to implement their approach.

Van der Veen \textit{et al}. \cite{van2015prioritizing} have implemented a pull request prioritisation tool for GitHub repositories to support integrators rather than code reviewers. They have also used different machine learning techniques, which also includes Random Forest, to implement their approach.

After reviewing the relevant related work, we identified the following limitations and research gaps in state of the art:

\begin{itemize}
    \item[$\cdot$] Code review requests have been prioritised considering the review requests per repository instead of considering the review requests per reviewer. Prioritising code review requests of different repositories in isolation provides an inefficient solution in the cases where reviewers have to review code from multiple repositories.
    \item[$\cdot$] It is often the case that existing approaches prioritise review requests based on the prediction of how fast the associated code changes are expected to be merged or with they will be merged at all. The problem of this type of approach is that it does not consider the nature of different code changes, which can be essential when it comes to defining what should be reviewed first (e.g., code changes that fix critical trouble reports).
    \item[$\cdot$] Given that software development involves a lot of uncertainty, it is essential to account for the existing uncertainty when prioritising code review requests. However, just one of the existing approaches used a technique that accounts for uncertainty and it does not do an end-to-end code review request prioritisation.
    \item[$\cdot$] To the best of our knowledge, none of the existing approaches have been evaluated in a live project in an industrial environment. Instead, existing approaches have been evaluated only through the use of historical data. As a consequence, there is no empirical evidence about the feasibility and usefulness of the existing prioritisation approaches in real and living software development undertakings.
\end{itemize}

In this paper, we address the aforementioned gaps by developing Pineapple. This tool uses a Bayesian network to prioritise code review requests accounting for the uncertainty inherent to software development. Furthermore, we evaluate Pineapple in a live, large-scale software development undertaking in Ericsson, focusing on predictive performance, feasibility, and usefulness.

\section{Research Design}
\label{chp:method}

In this section, we outline the design of our investigation. We have conducted an improvement case study \cite{runeson2009guidelines}. The case and unit of analysis is the project associated with the development of a telecommunication software product in Ericsson. The case project was selected through convenience sampling; it was a project with enough scale and where it was found the need to improve how code review requests are prioritised.

The development is carried out in three different centres of excellence located in Sweden, India, and Brazil. The 24 software teams involved in the case project use agile principles and practices (e.g., writing user stories, working in fixed-length sprints, stand-up meetings, and continuous integration). There are 5-7 developers in each team (153 software developers in total).

\begin{figure}[h!]
    \centering
    \includegraphics[width=\columnwidth]{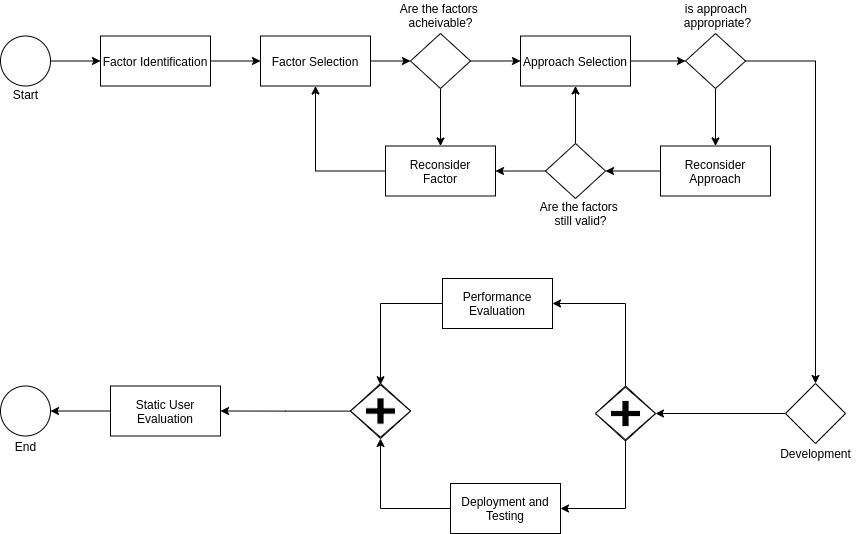}
    \caption{Research Design Structure}
    \label{fig:design}
\end{figure}

Figure \ref{fig:design} shows the steps of our investigation. \textbf{First}, to identify relevant factors to prioritise code review requests, we have conducted a literature review. To complement the results of the literature review, we have conducted semi-structured interviews, interviewing software developers involved with the case project.

\textbf{Second}, we have selected a subset of the factors identified in step 1. To systematize the factor selection process, we have defined the following criteria:

\begin{itemize}
    \item[\textbf{C1:}] The factor is present in the existing literature.
    \item[\textbf{C2:}] The factor is mentioned in the interviews.
    \item[\textbf{C3:}] The factor can be obtained using automated extraction.
    \item[\textbf{C4:}] The complexity of factor extraction is bearable.
    \item[\textbf{C5:}] The complexity of factor data analysis is bearable.
    \item[\textbf{C6:}] The obtained data is reliable.
\end{itemize}

\textbf{Third}, we have selected an approach to prioritise code review requests, considering both the nature of the problem at hand and the selected factors. To do so, we have surveyed existing literature to identify an approach that is suitable for the target problem and the available data.

\textbf{Fourth}, to evaluate our approach, we conducted a two-step approach:
\begin{itemize}
    \item \textbf{Predictive Performance Evaluation}~--~To answer RQ2, we have calculated the predictive performance of our approach using historical data. To do so, we calculated the metrics RMSE (Root Mean Square Error) \cite{chai2014root} and MAE (Mean Absolute Error) \cite{chai2014root}. Then, we compared the performance of Pineapple, which uses a Bayesian Network, against models implemented using the following Machine Learning techniques: Random Forest (Bagging), Gradient Boosting, Logistic Regression, and K-Nearest Neighbours.
    \item \textbf{Static Validation}~--~To answer RQ3, i.e., to evaluate the feasibility and usefulness of our approach in a live industrial environment, we conducted a static validation \cite{gorschek2006model}. It was operationalised in two steps. First, the users of Pineapple were asked to answer a questionnaire that was made available in the tool itself. In a second moment, we interviewed a subset of the questionnaire respondents to understand if their initial opinion about Pineapple changed after four months.
\end{itemize}

In the remainder of this section, we provide more details associated with the collection and analysis of the data used in our investigation.

\subsection{Data Collection and Analysis}

To collect the data required to answer our research questions, we employed the following methods: literature review, semi-structured interview, repository mining, and questionnaire.

To identify factors relevant to prioritise code review requests, we first conducted a \textbf{literature review}. The papers within the time frame of the last 20 years were considered in our literature survey. To survey the literature, we used the snowballing method. The start set for the snowballing method consisted of one paper \cite{van2015prioritizing}. We performed three iterations and selected 34 articles after completing both backward and forward snowballing.

To complement the literature review, we conducted \textbf{semi-structured interviews}. A total of 16 Ericsson software developers were interviewed. All the interviews were conducted face-to-face and took 30 minutes on average. We selected the interviewees using convenience sampling; we selected developers that often receive many code review requests. The interviewees' role and experience are presented in Table I of the online appendix\footnote{https://bit.ly/3jdP2nm}. The used interview guide is available in Table II of the online appendix.

During the interviews, we first asked the interviewees to look at the list of factors identified through the literature review and add any additional factor if necessary. After, the interviewees were asked to distribute 100 points among the listed factors. The result of this step was used in the factor selection step.

In a second moment, the interviewees were asked to draw on a paper the relationships between the factors (if any). They were asked to focus on how the factors impact each other. We used the result of this step to define the architecture of Pineapple's Bayesian Network.

After Pineapple was deployed in production, we conducted a 2-phase static validation. First, we have used a \textbf{questionnaire} (see Table III in the online appendix). The questionnaire was made available in Pineapple itself. We did so to allow the users to provide their opinion about the tool just after they used it. As a result, 24 software developers answered the questionnaire. We collected data during the first month of Pineapple's operation. Each user of Pineapple could answer the questionnaire just once. Furthermore, the users were not obliged to answer the questionnaire.

In the second phase of the static validation, we have conducted five additional semi-structured interviews to check if the opinion of the developers who participated in the first phase of Pineapple's static validation (through the questionnaire) changed after four months using the tool. To do so, we used the same aforementioned questionnaire used in the first phase of the static validation. We used the same sampling criteria used to select the first 16 interviewees. The interviewees' role and experience are presented in Table I of the online appendix.

To evaluate the predictive performance of Pineapple, we performed \textbf{repository mining} to extract data from the case project's repositories (110). 

Considering that Bayesian Networks produce a probability as output, the most adequate way to measure their predictive performance is through metrics that measure the rate of error and error margin. That is the reason we evaluated the predictive performance of Pineapple using RMSE (Equation \ref{for:rmse}) and MAE (Equation \ref{for:mae}). The values for both metrics vary from 0 to 1, where zero means the best performance possible.

\begin{equation}
    RMSE_{score} = \sqrt{\sum_{i=1}^{n}\frac{\left ( Predicted_{i} - Actual_{i}\right )^{2}}{n}}
\label{for:rmse}
\end{equation}

\begin{equation}
    MAE_{score} = \frac{1}{n}\sum_{i=1}^{n}\left | Actual_{i} - Predicted_{i} \right |
    \label{for:mae}
\end{equation}

To train Pineapple's Bayesian Network using the extracted data, it was necessary to discretise the data associated with the factors age, size, and the number of patches. The factors were discretised into three different categories using percentiles (see Table IV of the online appendix).

\section{Pineapple Overview}

An overview of how Pineapple prioritises code review requests is as follows:
\begin{itemize}
    \item Using a Bayesian Network, Pineapples calculates the merge probability of the code changes associated with each code review request assigned to a developer.
    \item The merge probability is used together with the change type (trouble report, feature, and product refactoring) and merge conflict (yes or no) factors to prioritise the existing review requests.
\end{itemize}

The reason we used merge probability (change status), merge conflict, and change type is that those were the factors deemed by the interviewees as the most relevant to prioritise code review requests.

The above-mentioned prioritisation approach is implemented using five microservices, which are deployed using Docker\footnote{www.docker.com} containers; each microservice has its own container. 

\begin{figure}[h!]
    \centering
    \includegraphics[width=\columnwidth]{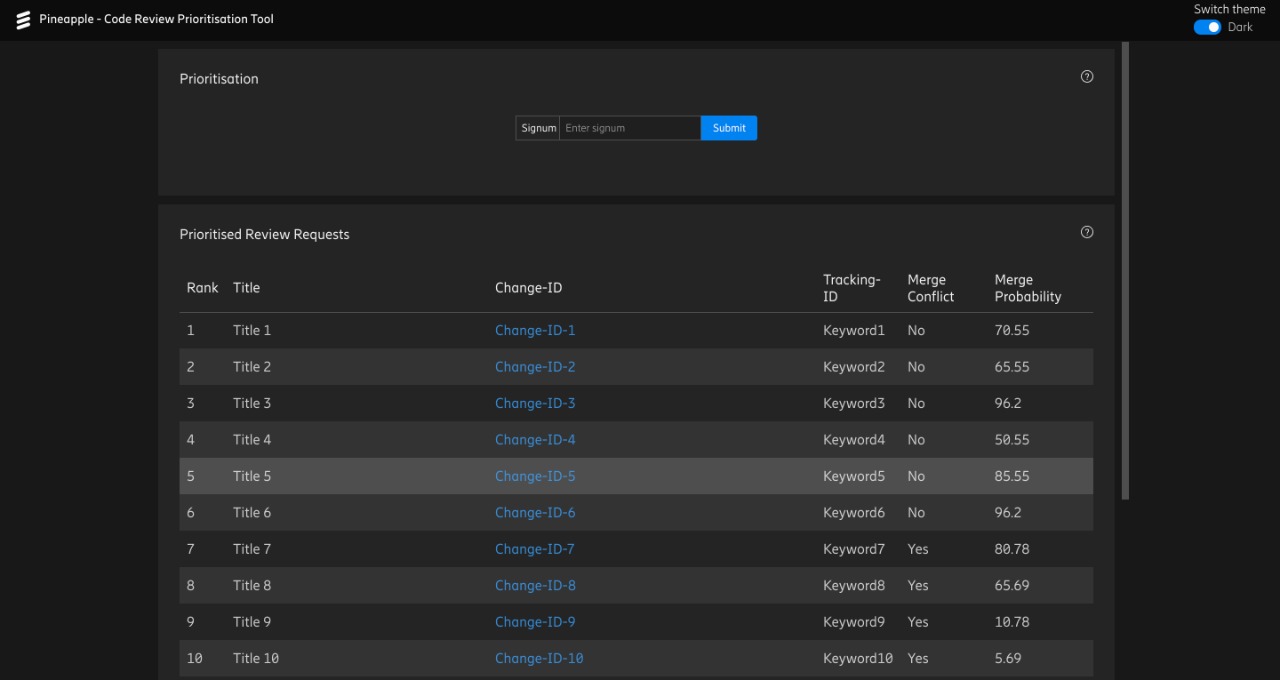}
    \caption{Pineapple Web Interface}
    \label{fig:pigui}
\end{figure}

Pineapple is a standalone tool. To use the tool (see Pineapples web interface in Figure \ref{fig:pigui}), a user needs to provide his/her user id. Then, Pineapple uses the provided id to fetch and prioritise the list of all open code review requests associated with the user. 

\begin{figure*}[h!]
    \centering
    \includegraphics[width=1.5\columnwidth]{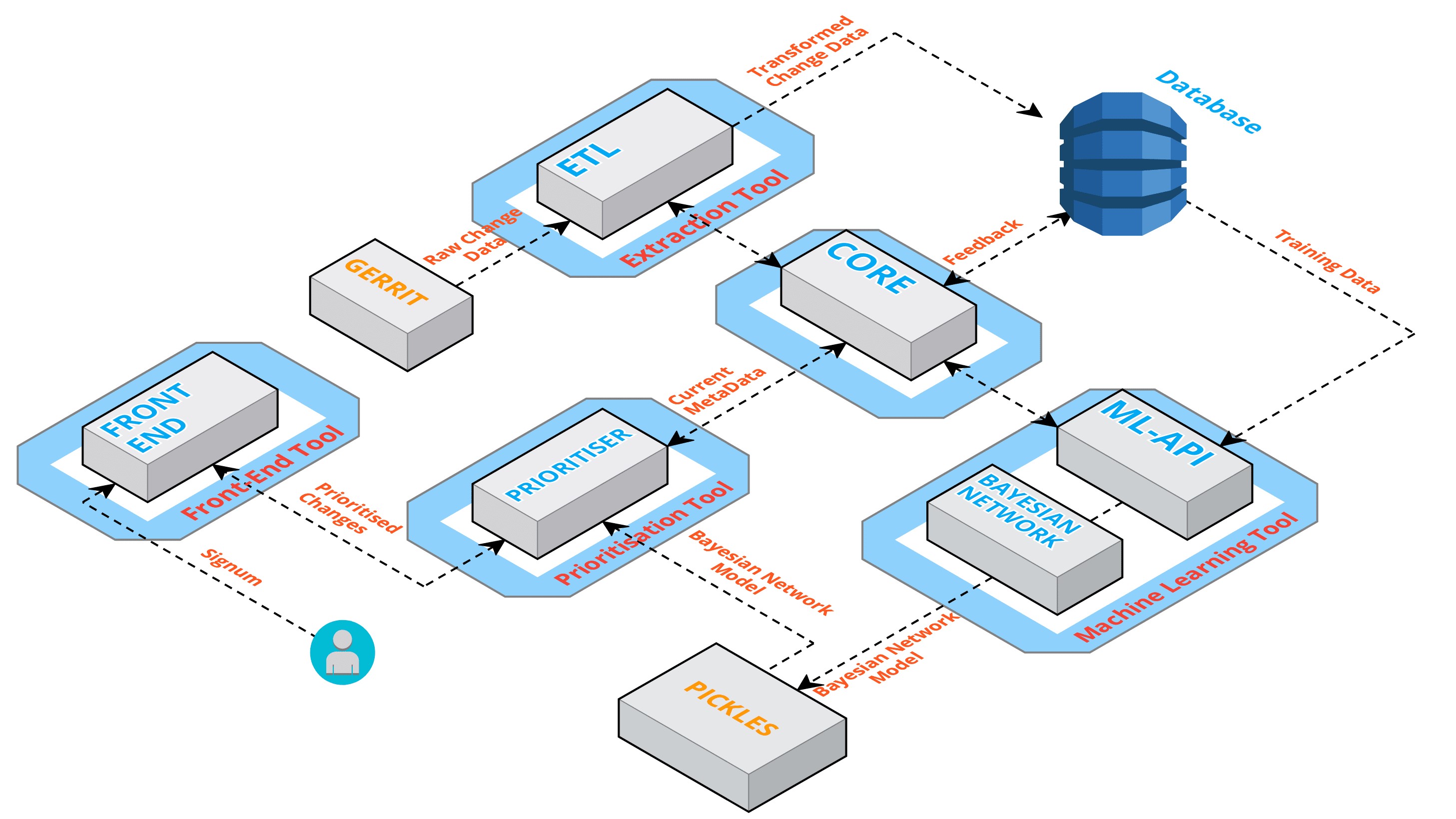}
    \caption{Pineapple Architecture Overview}
    \label{fig:paro}
\end{figure*}

The architecture of Pineapple is presented in Figure \ref{fig:paro}. The lines represent major data channels along with the direction and its response data. 

The five microservices are: Core, ETL, ML, Prioritiser, and Front-End. In the next subsections, we present more details about the services.

\subsection{Core, ETL, and Front-End Services}

The Core, ETL, and Front-End services are responsible for providing the basis for the training, serving, and monitoring of Pineapple's Bayesian Network (ML service), along with the generation of a list of prioritized review requests (Prioritiser service).

The \textbf{Core} service acts as the backbone for Pineapple. It acts as a messenger to communicate and bind all the services together. All the data and message transfer between the services is done through the Core. It also handles the scheduler, which triggers the ETL service to perform their tasks daily automatically. 

An additional task performed by the Core service is to store the responses provided by the users to the static validation questionnaire in a database. This was performed by the Core only during the static validation period.

The \textbf{ETL} service is responsible for \textbf{e}xtracting the data from the target version control/code review tool, \textbf{t}ransforming the raw data according to the needs of the Bayesian Network (e.g., data discretisation), and \textbf{l}oading the data in a database. In its current version, Pineapple is only able to extract data from Gerrit\footnote{www.gerritcodereview.com} repositories. The binning values used to discretise the factors age, size, and the number of patches are stored in a pickle\footnote{docs.python.org/3/library/pickle.html} file.

This service is configured to extract data from Gerrit repositories of the case project once per day. The ETL service is also responsible for extracting the user meta-data when the user provides his/her user id as input to prioritise his/her code review requests.

The \textbf{Front-End} service provides a web-based interface for Pineapple. Note that Pineapple can be used through REST API calls, which we plan to do to allow a smooth integration between Gerrit and Pineapple.

\subsection{ML Service}
\label{sec:bn}
The ML service uses the data extracted, transformed, and loaded by the ETL service to train Pineapple's Bayesian Network. The resulting model is stored in a pickle file. 

Pineapple retrains its Bayesian Network once per week. After retraining, Pineapple replaces the previous model by overwriting the existing pickle file.

In the remainder of this subsection, we present the factors we identified and selected to build Pineapple's Bayesian Network, along with the network architecture.

\subsubsection{Bayesian Network Factors}
\label{sec:rev_fac}
To identify and select the relevant factors to prioritize code review requests (RQ1), we first surveyed existing literature, which was complemented by semi-structure interviews. The results are presented in Table \ref{tab:fact}. 

We identified that the size, age, and type of a code change, along with automated tests' and peer reviews' verdicts, and merge conflicts are the most frequent factors. Code change type and automated tests were ranked the highest by the interviewees. 

\begin{table*}[h!]
\renewcommand{\arraystretch}{1.4}
\caption{Factors identified from Literature Review and Interviews}
\begin{adjustbox}{width=2\columnwidth,center}
\begin{tabular}{llllll}
\hline
{\color[HTML]{000000} \textbf{Factor}}     & {\color[HTML]{000000} \textbf{Description}}                                                                                      & \textbf{\begin{tabular}[c]{@{}l@{}}Int.\\ Freq.\end{tabular}} & \textbf{\begin{tabular}[c]{@{}l@{}}Lit.\\ Freq.\end{tabular}} & \textbf{\begin{tabular}[c]{@{}l@{}}Int.\\ Points\end{tabular}} & {\color[HTML]{000000} \textbf{\begin{tabular}[c]{@{}l@{}}Literature\\ Source\end{tabular}}} \\ \hline
Size of the change  & Number of add, del lines of code & 13                                                                     & 4                                                                       & 140                                                                 & \cite{van2015prioritizing,gousios2015work,zhao2019improving,fan2018early}  \\
Files changed      & Number of files added or deleted                                                                         & 1                                                                      & 4                                                                       & 8                                                                   & \cite{van2015prioritizing,gousios2015work,zhao2019improving,fan2018early}                                                       \\
Age          & Age of the change                                                                                     & 11                                                                     & 2                                                                       & 135                                                                 & \cite{van2015prioritizing,gousios2015work}                                                                 \\
Code Change Type  & Trouble report fixing,  & 16                                                                     & 2                                                                       & 544                                                                 & \cite{van2015prioritizing,gousios2015work}                                                              \\ & new feature, or refactoring \\
Test Verdicts              & The verdict of automated tests               & 16                                                                     & 3                                                                       & 322                                                                 & \cite{van2015prioritizing,gousios2015work,zhao2019improving}                                                             \\  &  tools \\
Developer stats     & Contribution and acceptance rate  & 5                                                                      & 4                                                                       & 72                                                                  & \cite{van2015prioritizing,gousios2015work,zhao2019improving,fan2018early}                                                             \\   &  along with activeness \\
Repository stats    & activeness of repository                                                                                             & 6                                                                      & 1                                                                       & 83                                                                  &  \cite{fan2018early}                                                                  \\
Number of revisions &  Number of patches in a change   request                                                             & 9                                                                      & 3                                                                       & 86                                                                  & \cite{van2015prioritizing,gousios2015work,zhao2019improving}                                                             \\
Peer Review  Verdicts                              & The verdict given by other reviewers   & 10                                                                     & -                                                                       & 147                                                                 & \hspace{5pt}-                                                                                            \\  &   \\ 
Merge Conflicts                            & Code conflicts faced while merging                                    & 16                                                                     & -                                                                       & 160                                                                 & \\
\hline
\end{tabular}
\end{adjustbox}
\label{tab:fact}
\end{table*}

Once we identified the factors and calculated their frequency and importance from the point of view of the interviewees, we applied the selection criteria presented in Section \ref{chp:method}. Table \ref{tab:fact_cri} shows a map between the identified factors and the specified criteria; an 'x' means that a factor fulfill a given criterion.
\begin{table}[h!]
\setlength{\tabcolsep}{2pt}
\caption{Factor Selection Criteria}
\begin{adjustbox}{width=0.8\columnwidth,center}
\begin{tabular}{llllll}
\hline
{\color[HTML]{000000} \textbf{Factor}}     & {\color[HTML]{000000} \textbf{C1}} & \textbf{C2} & \textbf{C3} & \textbf{C4} & {\color[HTML]{000000} \textbf{C5}} \\ \hline
{\color[HTML]{000000} Size of the change}  & {\color[HTML]{000000} x}           & x           & x           & x           & {\color[HTML]{000000} x}           \\
{\color[HTML]{000000} Files changed}       & {\color[HTML]{000000} x}           & x           & x           & x           & {\color[HTML]{000000} x}           \\
{\color[HTML]{000000} Age}                 & {\color[HTML]{000000} x}           & x           & x           & x           & {\color[HTML]{000000} x}           \\
{\color[HTML]{000000} Contain Keywords}    & {\color[HTML]{000000} x}           &             & x           &             & {\color[HTML]{000000} x}           \\
{\color[HTML]{000000} Tests}               & {\color[HTML]{000000} x}           & x           & x           & x           & {\color[HTML]{000000} x}           \\
{\color[HTML]{000000} Developer stats}     & {\color[HTML]{000000} x}           &             &            & x           & {\color[HTML]{000000} x}           \\
{\color[HTML]{000000} Repository stats}    & {\color[HTML]{000000} x}           &             & x           & x           & {\color[HTML]{000000} x}           \\
{\color[HTML]{000000} Number of revisions} & {\color[HTML]{000000} x}           & x           & x           & x           & {\color[HTML]{000000} x}           \\
Peer Review                                & x                                  & x           & x           & x           & x                                  \\
Merge Conflicts                            & x                                  & x           & x           & x           & x      \\
\hline                           
\end{tabular}
\end{adjustbox}
\label{tab:fact_cri}
\end{table}

The following is the motivation for the selected factors:
\begin{itemize}
    \item \textbf{Change type}~--~ All the 16 interviewees have mentioned that the type of change is significant and is one of the first things they consider when they review code review requests. Changes that fix trouble reports have higher priority than changes that add new features or changes related to refactoring of the product. This factor is not directly implemented in the Bayesian Network but is used by the Prioritizer service (see next subsection). The change type is extracted from the commit messages associated with code changes and can assume three values (or a combination of them): trouble report, new feature, or refactoring.

    \item \textbf{Test verdicts}~--~The existence of automated tests has been considered essential because the verdicts provided by these tests indicate if a given code change has problems that should be fixed before a code review session. The test verdicts also suggest if there is a need to focus on a particular change. This factor can assume the following values: -1 (failed), 0 (not finished), or 1 (passed).

    \item \textbf{Size}~--~Literature shows that large code changes are often defect-prone and have high chances of failing the automated tests. Another important aspect is that developers prefer to review small changes. This factor can assume the following values: Small, Medium, or Large

    \item \textbf{Age}~--~The state of the art shows that old code changes are often abandoned. Furthermore, the age of a code change relates to the number of its revisions/corrections; the older, the more revisions a code change tends to have. This factor is measured in minutes. The starting point for measuring age is when a review request is created (when the code is pushed for review for the first time), while the ending point is when Pineapple is requested to carry out prioritisation.

    \item \textbf{Peer Review Verdicts}~--~When a developer reviews a change request, she/he needs to know if the request has already been reviewed by another developer. Although it is often beneficial to have more than one reviewer per code change, it might not be worth to review a code change that has already received a negative feedback. Instead, it might be better to wait for a new version of the code change, including the expected corrections. Thus, this factor represents the existing verdicts associated with a given code change. It can have the following values: -2 (major corrections are required), -1 (minor corrections are required), 0 (no verdict), +1 (approved code change, but the approval from another reviewer is required), +2 (approved code change).

    \item \textbf{Number of Revisions}~--~Necessary corrections to code changes are reflected in the code through revisions (also known as patches). The presence of many revisions in a given code change may indicate that the change is problematic and, thus, should go through careful review. This factor can assume the following values: Low, Medium, or High.

    \item \textbf{Merge Conflicts}~--~Changes with merge conflicts require additional work before its code is ready for review; changes with merge conflicts cannot be merged. This factor is not included in Pineapple's Bayesian Network. Instead, it is used by the Prioritizer service together with change type and the merge probability calculated by the Bayesian Network. It can assume the following values: Yes or No.
\end{itemize}

\subsubsection{Bayesian Network Architecture}

A Bayesian Network consists of nodes connected to each other based on dependencies. The edges between nodes consist of the conditional probability distribution among the nodes. In general, a Bayesian Network has a terminal node, whose probability is used for decision making. 

\begin{figure}[h!]
    \centering
    \includegraphics[width=\columnwidth]{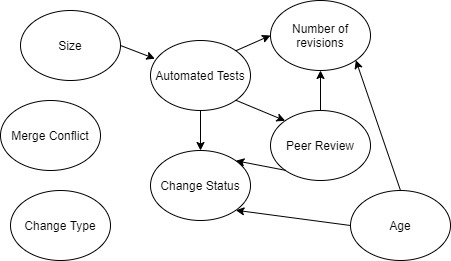}
    \caption{Bayesian Network Architecture}
    \label{fig:bn}
\end{figure}

The final architecture of Pineapple's Bayesian Network is presented in Figure \ref{fig:bn}.

The code \textbf{change status} (terminal) node represents the merge probability of a change. The \textbf{age}, \textbf{size}, represent respectively how old and big is a code change. The \textbf{number of patches} represents the number of patches/corrections associated with a given code change.

The nodes \textbf{merge-conflict} (it represents whether or not there is a conflict in a given code change) and \textbf{change-type} (it represents the task type associated with the code change) are independent variables which are not part of the Bayesian Network. They are used to sort the changes upon their importance, i.e., a change with merge conflict will be given lower rank compared with its counterpart. Similarly, the type of change also impacts the significance of the review request (a trouble report is given higher priority than any of the other change types).

\subsection{Prioritiser}

The Prioritiser first requests the ETL service (through the Core service) to extract meta-data (open code review requests) associated with the developer using the tool.

In a second moment, the Prioritiser service loads the trained Bayesian Network from a pickle file and calculates the merge probability of all code changes associated with the code review requests fetched in the first step. 

Finally, the Prioritizer uses the calculated merge probabilities, together with merge conflict and code change type factors, to prioritise a software developer's code review requests. The prioritisation is done as follows:
\begin{enumerate}
    \item The review requests are grouped based on the presence or absence of merge conflicts; requests with no merge conflicts have higher priority.
    \item The review requests in each of the aforementioned groups (no merge conflict and merge conflict) are further grouped based on the type of code change (trouble report fixing, new feature, refactoring). Pineapple first gives priority to trouble report fixing, second new features, and third refactoring.
    \item The final step is to prioritise the code review requests in each subgroup based on the calculated merge probability; the higher the merge probability, the higher the priority.
\end{enumerate}

The final result is a list with prioritized code review requests, which is presented by the Front-end service to the user of the tool (see Figure \ref{fig:pigui}).

\section{Evaluation Results and Analysis}
\label{chp:results}

In this section, we present the results of Pineapple'äs evaluation. We have organized this section around RQ2 and RQ3, which relate to the predictive performance evaluation and static validation (feasibility and usefulness) respectively. 

\subsection{RQ2 - Predictive Performance Evaluation}
\label{sc:mod_evl}
To evaluate Pineapple's predictive performance, we have used historical data extracted from the repositories of the case project. Furthermore, we implemented baseline models using the following machine learning techniques: Random Forest, Gradient Boosting, Logistic Regression, and K-Nearest Neighbours. We selected these approaches to be used as a baseline due to their popularity in both research and practice.

We selected a Bayesian approach because of its capability of handling uncertainty through probabilities (merge probabilities), i.e., Pineapple produces and uses probabilities to prioritize code review requests. It would not be desirable to round the probability values and convert Pineapple into a classifier; this would waste the advantage of Bayesian network related to handling uncertainty. Thus, to measure Pineapple's predictive performance, it was necessary to employ metrics used for measuring regression models.

We used both Root Mean Square Error (RMSE) and Mean Average Error (MAE) metrics to evaluate Pineapple's predictive performance and compare it with the baseline models. The primary purpose of these metrics is to know how much deviation is observed in the predicted values from the actual values. The metrics were calculated by performing k-fold cross-validation (5 folds), which splits the data randomly. 

Table \ref{tab:metrics} shows the obtained RMSE and MAE for Pineapple's Bayesian Network and the baseline models, along with the percentage difference between Pineapple and the baselines. When contrasting the performance of Pineapple's Bayesian Network against the baseline models (with the same historical data from the case project's repositories), Pineapple presented the best result regarding both RMSE (0.21, 24\% better than Gradient Boosting, the second-best) and MAE (0.15, 40\% better than Gradient Boosting, the second-best). This indicates that the selection of a Bayesian approach for Pineapple was adequate.

\begin{table}[h!]
\renewcommand{\arraystretch}{1.2}
\caption{Comparison of Model Evaluation Metrics}
\begin{adjustbox}{width=1\columnwidth,center}
\begin{tabular}{lll}
\hline
\textbf{Model}      & \textbf{RMSE} & \textbf{MAE} \\ \hline
Bayesian Network    & 0.21          & 0.15         \\
Random Forest       & 0.30 (+43\%)          & 0.25 (+67\%)          \\
Gradient Boosting             & 0.26 (+24\%)           & 0.21 (+40\%)         \\
Logistic Regression & 0.29 (+38\%)          & 0.23  (+53\%)        \\
K-Nearest Neighbours       & 0.31  (+48\%)         & 0.26   (+73\%)      \\\hline
\end{tabular}
\end{adjustbox}
\label{tab:metrics}
\end{table}

\subsection{RQ3 - Static Validation}
\label{sec:static_eval}

\begin{table*}[h!]
    \centering
    \caption{Developers' perceptions regarding Pineapple's Prioritisation}
    \label{tab:rq2.3a}
    \begin{adjustbox}{width=2.07\columnwidth,center}
    \begin{tabular}{lccccc}
    \hline
    Survey question & Strongly agree & Agree & Neutral & Disagree & Strongly disagree\\
    \hline
    \multicolumn{6}{c}{}\\
    Pineapple is useful to prioritise review requests & 21.7\% & 60.9\% & 4.3\% & 8.7\% & 4.3\% \\
    Pineapple is easy to use and understand & 39.1\% & 56.5\% & 4.3\% & 0\% & 0\% \\
    Pineapple helps in decreasing the lead time of review process & 4.3\% & 52.2\% & 26.1\% & 13.0\% & 4.3\% \\
    Pineapple  helps  inexperience  developers  in  performing  code reviews & 13.0\% & 34.8\% & 26.1\% & 21.7\% & 4.3\% \\
    Prioritisations done by Pineapple are reliable & 13.0\% & 47.8\% & 17.4\% & 17.4\% & 4.3\% \\
    \hline
    \multicolumn{6}{l}{Percentages in the table are rounded off to the nearest ten.}
    \end{tabular}
    \end{adjustbox}
\label{tab:evaluation}
\end{table*}

To evaluate the feasibility and usefulness of Pineapple in a large scale industrial environment, we conducted a static validation, i.e., these two aspects were evaluated based on the opinion of Pineapple's users. 

The first part of the static validation was operationalized through a questionnaire, which was made available in the tool itself for one month. As a result, 24 four unique Pineapple users (all software developers) answered the questionnaire.

Table \ref{tab:evaluation} shows the results related to the questions of the questionnaire. First,  Pineapple users were asked to answer, using a 5-point Likert scale, how useful are the prioritisation results produced by Pineapple. As a result, 61\% of the users agree, and 22\% strongly agree that Pineapple results are useful (83\% overall agreement level). The results indicate that users are satisfied with the tool and believe it can help them with code review request prioritisation. 

Second, the users were asked to answer about the usability of Pineapple using a 5-point Likert scale. As a result, 57\% of the users agree, and 39\%  strongly agree (96\% overall agreement) that Pineapple is easy to use and understand.

Third, we asked the users to respond, using a 5-point Likert scale, about the extent to which Pineapple can help reducing code review lead time. The results show that the majority of the developers either agree (57, 53\% agree and 4\% strongly agree) or are neutral (26\%), and only 17\% disagree that Pineapple can help reducing review lead time. Overall, this is also a positive result for Pineapple.

When it comes to the extent to which Pineapple can help new developers in the case project regarding code reviews (fourth question), the majority of the respondents either are neutral (26\%) or disagree (26\%, 22\% agree, 4\% strongly disagree), while 48\% agree that Pineapple can help new developers (35\% agree, 13\% strongly agree). This indicates that there is no clear consensus about the topic aimed by the question.

The final question of the questionnaire is about the reliability of Pineapple results. The users were asked to answer this question using a 5-point scale. As a result, 61\% of the respondents (48\% agree and 13\% strongly agree) agree that the prioritisation results produced by Pineapple are reliable, while 17\% are neutral and only 22\% disagree (18\% disagree and 4\% strongly disagree). This indicates that most developers are satisfied about our tool's accuracy.

Considering that we obtained the aforementioned results in the first month of operation of Pineapple, we decided to check the opinion of Pineapple's users again after some time (four months), to see if there was any change in the users' feelings towards our tool. To do so, we interviewed five developers that have continuously used Pineapple. 

Through the results of the five interviews, we learned that all five interviewees were still very positive regarding the feasibility of Pineapple, especially regarding its predictive performance. Four of them were also still very positive regarding Pineapple's usefulness. The remaining developer was neutral. He justified that, in his specific case, he did not have enough code review requests assigned to him to motivate a frequent use of the tool. However, he had used the tool eventually and continued to like it.

\section{Discussion}
\label{chp:discussion}

Code review request prioritisation is not a topic widely addressed by the software engineering community \cite{Badampudi:2019}. In addition to that, there was no empirical evidence of the usefulness and feasibility of this type of approach in real live industrial environments, since all existing studies have only evaluated their respective proposals from a predictive performance point of view (through historical data).

To address the gaps identified in state of the art, we have developed Pineapple using a Bayesian approach, which is known for handling uncertainty in a very effective way \cite{Hamilton:2012}. We have also evaluated it in a real live project.

All the related approaches in literature handle the prioritisation problem indirectly, by classifying code review requests in different categories (e.g., whether or not a given change will be merged). Considering that Pineapple does the end-to-end prioritisation and produces a numerical number as the result of its Bayesian Network, it is not easy to compare our approach with the ones in the existing literature.

\begin{figure}[h!]
    \centering
    \includegraphics[width=\columnwidth]{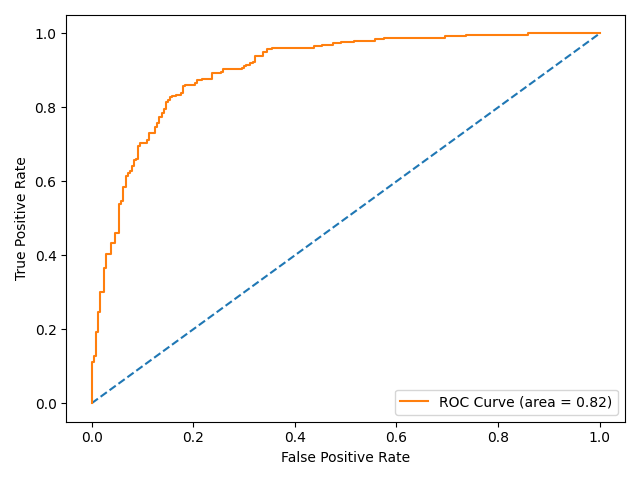}
    \caption{ROC Curve}
    \label{fig:roc_curve}
\end{figure}

Given the aforementioned challenge, we made an attempt to compare the predictive performance of Pineapples Bayesian Network with the approach proposed by Fan \textit{et al.} \cite{fan2018early} (Random Forest-based), which is the one that is the most similar to Pineapple. To do so, we rounded the probabilities generated by Pineapple and calculated the Area Under the Curve (AUC) metric to compare with the value reported by Fan \textit{et al.}. As a result, we found that Pineapple's AUC (0.82) is superior to the one reported by Fan \textit{et al.} (0.74). Although Pineapple was developed to use probabilities, it still performs well when the predicted probabilities are rounded to use Pineapple as a classifier. Figure \ref{fig:roc_curve} represents the ROC curve of Pineapple's Bayesian Network.

Another interesting finding is that despite being a very popular machine learning approach in the literature, Random Forest presented a performance that was inferior to our Bayesian-based approach both in the regression (see Section \ref{sc:mod_evl}) and in the classification scenarios (see the results mention earlier in this section). Random Forest also performed worse than two of the other approaches we compared with Pineapple in the regression scenario (Gradient Boosting and Logistic Regression models, see Section \ref{sc:mod_evl}).

Another relevant contribution of this paper is the conducted static validation of Pineapple. In doing so, we made an attempt to address the lack of empirical evidence regarding the usefulness and feasibility of code review request prioritisation approaches based on machine intelligence. While we acknowledge that the scale of the conducted evaluation is not enough to advocate for the use of this type of tool in industrial environments, it helped us to do a sanity check and understand the value of this type of approach from the point of developers in our organisation. That said, we believe it is worth investigating if the factors included in Pineapple fit other contexts as well; it may be necessary to adapt the list of factors in other scenarios.

The employed research design allowed us to introduce the developers' knowledge into Pineapple early on. While conducting the interviews, we observed that the majority of the software developers were very interested in our investigation and wanted to use the tool whenever available. 

We believe that this paper contributes both to the states of the art and practice by presenting an end-to-end prioritisation approach that accounts for uncertainty by design. Furthermore, the evaluation approach used here can be used by researchers to improve the relevance of their approaches. Finally, the microservice-based architecture of Pineapple addresses practical challenges associated with serving and monitoring machine intelligence-based models. Hence, it can be used as inspiration by practitioners seeking to use machine intelligence to improve software development efficiency and effectiveness. 

\section{Threats to Validity}

We have followed Runeson's guidelines \cite{runeson2009guidelines} and discussed the threats to the validity of our investigation in terms of reliability, construct validity, internal validity, and external validity.

\textbf{Reliability} concerns the extent to which it is possible to reproduce the results of an investigation \cite{runeson2009guidelines}. A significant part of the data used in our investigation is qualitative and was obtained through interviews. The threats to the validity of our investigation in this category include interviewee and interviewer bias. To mitigate these threats, the two authors developed a case study protocol together to select interviewees and systematically conduct the interviews, i.e., it includes the interview guide and questionnaire used in our investigation. While the interviews were conducted only by the first author, the second author reviewed the raw data and analyzed the results. One reliability limitation of this study is the fact that we are not able to share with the research community the data used in our investigation given its sensitivity for the case company.

\textbf{Construct} validity concerns the extent to which the operational measures used in our investigation represent and answer our research questions \cite{runeson2009guidelines}. In this category, there are two main threats to the validity of our investigation: misinterpretation of the interview questions and the questionnaire questions. To mitigate the first threat, at the start of the interviews, we clarified the purpose and how the results would be used. Regarding the questionnaire, we used an iterative process to design and improve it, aiming at avoiding ambiguities.

\textbf{Internal} validity relates to the confounding factors that could impact the validity of our results \cite{runeson2009guidelines}. It may be possible that we have not included all factors necessary to prioritize code review requests in the best way possible. To mitigate this threat, we have used method triangulation (literature review and semi-structured interview). Another validity threat relates to the extent to which the evaluation results relate to the introduction of Pineapple in the case project or are due to confounding factors not controlled in our investigation. To mitigate this threat, we also use method triangulation (semi-structured interview and questionnaire). However, we acknowledge that it would be necessary to evaluate the tool in a semi-controlled environment, account for potential confounding factors, and use project data to evaluate Pineapple's usefulness and feasibility. In doing so, we could increase the significance of the evaluation results. 

\textbf{External} validity concerns the extent to which the findings can be generalized and are interesting outside the investigated case \cite{runeson2009guidelines}. Our results have limited generalizability; we have evaluated Pineapple in just one case (one project and one company). Nevertheless, we believe our results can indicate the feasibility and usefulness of code review request prioritization tools in industrial environments.  

\section{Conclusions and Future Work}
In this paper, we conducted an improvement case study that resulted in Pineapple, a code review request prioritisation tool. We have used both the relevant literature and expert knowledge from Ericsson developers to develop the tool. Furthermore, we have evaluated the tool in a live real large scale industrial environment.

To answer RQ1, we have conducted a literature review and semi-structured interviews. We used the results to build Pineapple's Bayesian Network, which produces one of the inputs for its prioritiser service.

To evaluate Pineapple's predictive performance (RQ2), we used historical data from the case project. We calculated RMSE and MAE and compared Pineapple's Bayesian Network with baseline models (implemented using state-of-the-art Machine Learning techniques). The results show that Pineapple has a better predictive performance than the baseline models.

We also evaluated Pineapple's feasibility and usefulness. To do so, we conducted a static validation to obtain the opinion of users about Pineapple. We used a questionnaire and conducted semi-structure interviews. The overall conclusion is that the results produced by Pineapple are considered reliable by the majority of its users and the tool is considered easy to use. Furthermore, the results are considered useful for prioritising code review requests and can help to reduce code review lead times. Our overall conclusion is that Pineapple is useful, feasible, and reliable. 

While the current evaluation results indicate that it is worth using Pineapple to prioritise code review requests, we believe it is still necessary to evaluate further the extent to which the produced prioritisation results are accurate enough and whether or not it contributes to shortening code review lead times. We plan to carry out a longitudinal investigation to address the aforementioned questions.

\bibliography{thesis-refs}
\bibliographystyle{unsrt}

\end{document}